\documentclass[preprint,showpacs,preprintnumbers,amsmath,amssymb]{revtex4}

\usepackage{graphicx}% Include figure files
\usepackage{dcolumn}% Align table columns on decimal point
\usepackage{bm}% bold math

\begin{document}

\title{The additive generalization of the Boltzmann entropy} 

\author{Alexander N. Gorban}
\email{gorban@icm.krasn.ru}
\affiliation{ETH Z\"{u}rich, Department of Materials, Institute of Polymers, 
ETH-Zentrum,  CH-8092 Z\"{u}rich, Switzerland\\
Institute of Computational Modeling RAS, 660036 Krasnoyarsk, Russia}

\author{Iliya V. Karlin}
%\homepage{http://mat.ethz.ch/d-werk/oettinger/MK_DIR/.people/iliya.html}
\altaffiliation[]{Corresponding author}
\email{ikarlin@ifp.mat.ethz.ch}
\affiliation{
ETH Z\"{u}rich, Department of Materials, Institute of Polymers, 
ETH-Zentrum,  CH-8092 Z\"{u}rich, 
Switzerland
}

\author{Hans Christian \"Ottinger}
\email{hco@ifp.mat.ethz.ch}
\affiliation{
ETH Z\"{u}rich, Department of Materials, Institute of Polymers,
ETH-Zentrum, CH-8092 Z\"{u}rich, 
Switzerland
}

\date{\today}

\begin{abstract}
There exists  only one generalization of the classical 
Boltzmann-Gibbs-Shannon entropy functional to a one-parametric family
of additive entropy functionals. We find analytical solution to the corresponding extension
of the classical ensembles, and discuss in some detail the example of the
deformation of the uncorrelated state.

%In this paper, we demonstrate that, 
%for a recently found family of non-classical  additive entropy functions 
%of Markov processes, the entropy maximum problem 
%is solved with the same degree of
%explicitness as for the classical .

\end{abstract}

\maketitle

\section{Introduction}\label{intro}

The growing interest to non-classical entropies in recent years 
 \cite{Tsallis88,Tpage} is motivated by the fact that they can be used to describe observable 
statistical effects such as:
(i) Non-classical tails of distribution functions which can deviate  significantly from 
Gaussian  distribution. In particular, this asymptotics can be 
power-law (``long tails'') or, instead, distribution functions 
can decay in a more rapid fashion (``short tails''), 
in particular, 
they can become equal to zero at finite distance (``cut tails'').
(ii). Strong correlations between subsystems in equilibrium and conditionally-equilibrium
 (quasi-equilibrium) states.
(iii) In particular, even at a vanishingly weak interaction between subsystems, when the 
classical Boltzmann-Gibbs-Shannon entropy (BGS) would lead to no correlations, 
strong correlations can be observed in the equilibrium. 
This may sound somewhat paradoxal: Joining non-interacting subsystems with equal
values of the intensive quantities, and switching on an infinitesimal weak interaction,
we produce a strongly correlated equilibrium. However, the simplest example is readily provided
(though not related to non-classical entropies {\it per se}) by the microcanonical ensemble
of finite systems: If subsystems are not interacting at all, then there is an additional 
conservation law, 
the energies of the individual subsystem, and the product of the microcanonic
distributions is the equilibrium. However, an arbitrarily weak interaction will surely destroy this
conservation law, and the equilibrium becomes the usual microcanonic ensemble (the
equipartition over the surface of constant total energy). For finite number of particles
in the subsystems, this latter state is correlated,  and it does  {\it not} factor into the product
of the microcanonical distributions of the subsystems. It is only in the thermodynamic limit
where the theorem about the equivalence of the ensembles  \cite{Ruelle} 
states the tendency to zero
of correlations of (almost) noninteracting subsystems (in the domain of its applicability, 
of course).  
We should remark that empirically found asymptotics of the distribution functions 
should be always treated  with care since they can turn out to be ``intermediate asymptotics''
rather than true limits.

The entropic description of all these effects in the spirit of Gibbs ensembles 
 is technically advantageous (same as any variational
principle) but this is by far not the only merit. If the entropy is consistent with the kinetics, 
and varies monotonically in time, then a very useful construction becomes available.
This is the conditional equilibrium (or quasi-equilibrium, with local equilibrium as a specific
example). The quasi-equilibrium describes partially relaxed systems, according to the idea of
 the 
fast-slow decomposition of motions: Fast variables have almost reached equilibrium at
almost fixed values of slow variables. Conditional equilibrium is described as the probability
distribution which brings to maximum the entropy $S(p)$ at fixed values
of the slow variables, $M=m(p)$:
\begin{equation}
\label{smax} S(p)\to\ {\rm max},\ m(p)=M.
\end{equation}

Usually, when one attempts to introduce non-classical entropies in order to use these
advantages, there is a price to be paid. Non-classical entropies at use in most of 
the contemporary studies violate at least one of  the following important and familiar properties
of the BGS entropy:
(i) Additivity: The entropy of the system which is composed of  independent
subsystems equals the sum of the entropies of the subsystems.
(ii) Trace-form: The entropy is  a sum over the states (see below).
(iii) Concavity of the entropy.
For example, the Tsallis entropy \cite{Tsallis88} is not additive, the R\'enyi entropy \cite{Renyi70} 
is not of the trace form.

Violation of additivity cannot be motivated by the fact that ``in reality, all subsystems are interacting'' \cite{Beck02}.
Indeed, the additivity axiom is the conditional statement: {\it If} the systems are independent, 
{\it then} the entropy of the joint system equals the sum of the entropies of subsystems. 
Probability theory, even when studying whatever strongly dependent events, is based
on such notions as independence, independent trials etc \cite{Kac57}. 
Giving up these notions simply
on the grounds that events in nature depend on each other is misleading.

In this paper we demonstrate how the description of both long and short tail
distributions, growth of correlations etc can be achieved on the basis of the entropy approach, 
and
without a violation of neither the additivity nor  of the trace form requirements 
(however, with a violation of the
concavity only for the description of cut tail distributions). Such a description becomes available
only if one uses a one-parametric family of entropies introduced recently \cite{GK02_tsallis}.
We establish analytic formulae for conditional maximizers of these entropies which 
makes operations within the present formalism almost
as easy as in the case of a Gaussian distribution pertinent to the BGS entropy.

\section{Additive trace-form entropies for Markov processes}
\label{basic}
The basic model we consider here is the finite Markov chain 
(finiteness and discreteness are by no means
the crucial  restriction, and are employed only in order to avoid the convergence questions). 
The time evolution of
the probabilities $p_i$, where $i$ is the discrete label of the state, is given by master equation,

\begin{equation}
\label{Markov1}
\dot{p}_i=\sum_{j,j\ne i}k_{ij}\left(\frac{p_{j}}{p^*_{j}}-\frac{p_i}{p^*_i}\right),\ k_{ij}=k_{ji}\ge 0.
\end{equation}
We consider only systems which allow for a positive equilibrium, $p^*_i>0$ (for infinite
systems, it is often advantageous to use unnormalized $p^*$). We recall 
\cite{Gorban84,GK02_tsallis}
that, for each convex function of one variable, $h(x)$, one constructs the Lyapunov
function $H_h(p)$ which does not increase on solutions to Eq.\ (\ref{Markov1}), where
\begin{equation}
\label{Lyap1}
H_h(p)=\sum_{i} p_i^*h(p_i/p_i^*).
\end{equation}
[We consider below $H_h$-functions rather than entropy functions $S_h=-H_h$.]

Among the set of Lyapunov functions (\ref{Lyap1}), there exists a one-parametric subset
of additive Lyapunov functions, $H_{\alpha}$, $0\ge \alpha\ge 1$:
\begin{eqnarray}
H_{\alpha}&=&\sum_{i} p_i^*h_{\alpha}(p_i/p_i^*),\nonumber\\
h_{\alpha}(x)&=&(1-\alpha)x\ln x-\alpha\ln x.
\label{result1}
\end{eqnarray}
In particular, 
\begin{eqnarray*}
H_0&=&\sum_{i} p_i\ln(p_i/p_i^*),\\
H_1&=&-\sum_{i} p_i^*\ln(p_i/p_i^*).
\end{eqnarray*}
Additivity of functions $H_{\alpha}$ (\ref{result1}) is readily checked 
\cite{Gorban84,GK02_tsallis}:
If $p=p_{ij}=q_ir_j$, and also  if $p^*=p^*_{ij}=q^*_ir^*_j$, then
\[ H_{\alpha}(p)=H_{\alpha}(q)+H_{\alpha}(r).\]
It can be demonstrated that the family (\ref{result1}) is unique (up to a constant factor): 
There are no
other additive trace-form  functions among Lyapunov functions (\ref{Lyap1})  of master equation.

\section{Solution to the maximum entropy problem}
\label{analytic}

Since a factor in front of $H_{\alpha}$ is irrelevant, it
 proves convenient to use a different parameterization of the family (\ref{result1}),
\begin{equation}
H_{\alpha}=\sum_{i}[p_i\ln(p_i/p_i^*)-\alpha p_i^*\ln(p_i/p_i^*)],
\label{family3}
\end{equation}
where $\alpha\ge0$, and where the case $\alpha\to\infty$ 
should be considered separately,
\begin{equation}
H_{\infty}=-\sum_i p_i^*\ln(p/p_i^*).
\label{burg}
\end{equation}
In order to address the construction of the quasi-equilibrium in a general setting, 
we assume the macroscopic variables
$M=m(p)$, where $M_s=\sum_i m_{si}p_i$, and consider the problem (\ref{smax}) 
with $S=-H_{\alpha}$. 
Solving this problem with the method of Lagrange multipliers, we find:
\begin{equation}
\frac{\partial H_{\alpha}}{\partial p_i}=\lambda_0+\sum_s \lambda_sm_{si},
\label{grad}
\end{equation}
where Lagrange multiplier $\lambda_0$ corresponds to normalization, and $\lambda_s$ 
to the rest of the constraints.
Let us denote $-\Lambda_i$ the right hand side of Eq.\ (\ref{grad}). With this, Eq.\ (\ref{grad}) 
may be written,
\begin{equation}
\ln(p_i/p_i^*)-\alpha(p^*_i/p_i)=-\Lambda_i.
\end{equation}
Solution to an equation,
\begin{equation}
\ln q-\alpha q^{-1}=-\Lambda,
\end{equation}
may be written as follows:
\begin{equation}
\label{soln1}
q=e^{-\Lambda}e^{{\rm lm}\left(\alpha e^{\Lambda}\right)},
\end{equation}
where we have introduced notation ${\rm lm}a$ (modified logarithm)
 for the function which is the solution 
to the
transcendent equation,
\[ xe^x=a.\]
The function ${\rm lm}$ satisfies the following identities:
\begin{eqnarray}
{\rm lm} a&=&\ln a-\ln {\rm lm} a,\label{identity1}\\
{\rm lm} a&=&\ln a-\ln(\ln a-\ln(\ln a-\ln(\dots))\dots).\label{identity2}
\end{eqnarray}
Identity (\ref{identity2}) is the recurrent application of identity (\ref{identity1}).
A different representation of solution (\ref{soln1}) reads:
\begin{equation}
\label{soln2}
q=\frac{\alpha}{{\rm lm}(\alpha e^{\Lambda})}.
\end{equation}
From the representation (\ref{soln1}), the asymptotics at $\alpha\to0$, 
and fixed $\Lambda$, is obvious:
$q\to e^{-\Lambda}$, and which corresponds to the usual Boltzmann distribution.
On the other hand, representation (\ref{soln2}) reveals the asymptotics at $\Lambda\to\infty$:
\[ q\sim\frac{\alpha}{\ln\alpha +\Lambda}.\]
For a symmetric distribution on the axis, and for $\Lambda=\lambda_0+\lambda_2x^2$, 
the first of the limits just
mentioned gives the Gaussian distribution, while the second limit gives the Cauchy distribution.
The corresponding distribution function for the limiting case $H_{\infty}$ is simply the Cauchy
 distribution
on the axis.
Among non-symmetric Cauchy distributions of the form, 
$p=(\lambda_0+\lambda_1x+\lambda_2x^2)^{-1}$,
there are distinguished cases with a twice degenerated zero in the denominator:
$p=(\lambda(x-a))^{-2}$. When one attempts to normalize this distribution by choosing a 
convergent 
sequence of functions, one gets a Dirac $\delta$-function, $\delta(x-a)$ 
which can be interpreted as 
a microcanonic ensemble. 

Thus, the quasi-equilibrium distribution has the form:
\begin{equation}
\label{result2}
p=p^*e^{-\Lambda}e^{{\rm lm}(\alpha e^{\Lambda})}= \frac{\alpha p^*}{{\rm lm}(\alpha e^{\Lambda})}.
\end{equation}
[We have omitted indices of states in $p$, $p^*$, and $\Lambda$.] Formula (\ref{result2}) is the main
result of this paper.

It is also interesting to address the formal extension of the result (\ref{result2}) to negative $\alpha$.
Function ${\rm lm}a$ is defined and is continuous for $a\ge -e^{-1}$ 
(${\rm lm}a\ge-1$). At $a\to-e^{-1}$, we have
the limit, $d{\rm lm}a/da\to\infty$. If we formally extend, ${\rm lm}a=-\infty$ for $a<-e^{-1}$,
then Eq (\ref{result2}) is a distribution with ``cut tail''. With this, there will be defined a non-zero
ratio $p/p^*$:
\begin{equation}
{\rm inf}\{p/p^*|p\ne 0\}\ge |\alpha|>0,
\end{equation}
that is, either $p\ge|\alpha|p^*$, or $p=0$. This construction is similar to a Maxwell construction of 
a stretched spinodal (the cut at the inflection point), and not to the global maximum of the entropy.
Whereas such constructions are always necessary when working with non-convex thermodynamic
potentials,  will not further discuss the case $\alpha<0$ in this paper.

\section{Quasi-equilibrium ensembles near the BGS limit} 

For the classical BGS entropy ($\alpha=0$), the quasi-equilibrium distribution has the form:
\begin{equation}
\label{boltzmann}
p=p^*e^{-\Lambda},
\end{equation}
where $\Lambda$ is the corresponding gradient of the entropy at the quasi-equilibrium, 
expressed in terms
of Lagrange multipliers.
Let us study the quasi-equilibrium (\ref{result2}) for small $\alpha$. To the first order, we get:
\begin{equation}
\label{1order}
p=p^*\left(e^{-\Lambda}+\alpha\right)+o(\alpha).
\end{equation}
[Note that, in this expansion, dependence of $\Lambda$ on the values of
the macroscopic variables $M$ is implicit. Explicit evaluation of this dependence
requires, in addition, an expansion of $\Lambda$ in terms of $\alpha$ which is 
used below when studying concrete examples.]

Substituting equation (\ref{1order}) into the  kinetic equation (Markov chain in
the present context,  into the  Liouville equation in the context of particle's dynamics, or,
generally speaking, into the linear  equation of the microscopic Markovian process), 
we easily see that the term $\alpha p^*$
gives no contribution to the resulting quasi-equilibrium dynamics. 
Indeed, we first notice that the relation between
the time derivative $\dot{p}$  with the Lagrange multipliers is the same, as for the 
classical Boltzmann's distribution (\ref{boltzmann}): If $L$ is the linear operator of Markovian
dynamics, $\dot{p}=Lp$, then, substituting for $p$ on the right hand side of this 
equation the expression (\ref{1order}), and using linearity, we get 
\[L\left(p^*\left(e^{-\Lambda}+\alpha\right)\right)=L\left(p^*e^{-\Lambda}\right).\]
Furthermore, defining the shifted macroscopic variables, $M_{\alpha}=m(p-\alpha p^*)$, we find that
for  the classical quasi-equilibrium dynamic 
equation, $dM/dt=m(Lp(M))$, where $Lp(M)$ is the microscopic vector field evaluated
at the classical quasi-equilibrium states $p(M)=p^*\exp(-\Lambda(M))$ is affected only by a shift 
$M\to M_{\alpha}$,
to the first order in $\alpha$. In other words, the quasi-equilibrium dynamics of the ensemble 
(\ref{result2}) is driven by the
classical dynamics resulting from the BGS entropy and Boltzmann distributions
(\ref{boltzmann}) to the first order in $\alpha$.

In order to compute the quasi-equilibrium to second order in $\alpha$, we must use the expansion of
 ${\rm lm}a$
to third order,
\[ {\rm lm}a=a-a^2 +(3/2)a^3 +o(a^3).\]
Then
\begin{equation}
\label{2order}
p=p^*\left(e^{-\Lambda}+\alpha-\frac{1}{2}\alpha^2e^{\Lambda}\right)+o(\alpha^2).
\end{equation}
Further corrections can also be easily computed using higher-order terms in the expansion of
the ${\rm lm}$. We now shall consider a specific example of the formula (\ref{2order}).

\section{Example: Enhancement of particle's correlations}
\label{example}

In order to illustrate  the effect of second order deviations from the BGS case,
we apply Eq.\ (\ref{2order}) to the classical quasi-equilibrium defined by the 
one-particle configurational distribution function
$f_1(r)$, where $r$ is position variable.
Assuming, as usual,  the equipartition for the reference equilibrium, $p^*=1/V^N$, 
where $V$ is the volume
of the system, and $N$ is the number of particles, we get
$e^{-\Lambda}=e^{\lambda_0}\prod_{i=1}^N\Psi(r_i)$, where 
Lagrange multiplier $\lambda_0$ is responsible for normalization. Then
the $N$-body quasi-equilibrium distribution function to second order in $\alpha$ reads,
\begin{equation}
\label{QEN}
V^Np=e^{\lambda_0}\prod_{i=1}^N\Psi(r_i)+\alpha-\frac{\alpha^2}{2e^{\lambda_0}\prod_{i=1}^N\Psi(r_i)}+o(\alpha^2).
\end{equation}
Our goal now is to compute the two-body configurational distribution function,
\[f_2(r,q)=N(N-1)\int p(r,q,r_3,\dots, r_N)dr_3\dots dr_N,\]
in the quasi-equilibrium (\ref{QEN}). We recall that the classical result for the BGS entropy
gives the uncorrelated two-body distribution, $f_2(r,q)\sim f_1(r)f_1(q)$, which also corresponds to  
the limit
($\alpha=0$)  of Eq.\ (\ref{QEN}). Computation to the order $\alpha^2$ requires expansion of Lagrange
multipliers $\lambda_0$ and $\Psi$ to the corresponding order. 
This computation is straightforward although
tedious, thus we give here only the final result: 
The two-body quasi-equilibrium configurational distribution
function $f_2$ reads:

\begin{equation}
\frac{N}{N-1}f_2(r,q)=
(1+\alpha +\alpha^2)\tilde{f}_1(r)\tilde{f}_1(q)+\alpha n^2
-\underline{
\frac{\alpha^2}{2}n^2B^N\varphi_1(r)\varphi_1(q)
} 
+o(\alpha^2), \label{RESULT}
\end{equation}
where $n=N/V$ is the average number density, and where we have introduced notation, 
\begin{eqnarray}
\tilde{f}_1(r)&=&f_1(r)-\alpha n,\label{shift}\\
\varphi_1(r)&=&\frac{f_1(r)}{n}-\frac{n}{Bf_1(r)},\label{holes}\\
B&=&\frac{1}{V}\int_{V}\frac{n}{f_1(r)}dr. \label{B}
\end{eqnarray}
It is readily checked that the result (\ref{RESULT}) gives $f_2=(N-1)N^{-1}f_1f_1$ at $\alpha=0$ which
is identical with the classical uncorrelated pair distribution with correct normalization \cite{BGK91}.

The first two terms in Eq.\ (\ref{RESULT})
amount again  to the  uncorrelated state with {\it homogeneously shifted} one-particle distributions
($\tilde{f}_1$ (\ref{shift}) instead of $f_1$, which 
 amounts to a homogeneous subtraction of the average density times $\alpha$).
The underlined term  (of the order of $\alpha^2$), is the contribution
responsible for correlations due to
the use of the non-classical entropy. 
It also has a form of a product, but not of the distribution functions, rather, of 
functions of one variable  (\ref{holes}). 
In order to see the effect of this term  more explicitly, we assume 
\begin{equation}
f_1(r)=n(1+\zeta(r)N^{-1/2}),
\end{equation}
where $\zeta$ is a function with zero average, and finite amplitude, $\langle\zeta\rangle=0$,
$\langle\zeta^2\rangle=\sigma^2$, where we have introduced notation for averaging over the volume,
$\langle h\rangle=V^{-1}\int_{V} hdr$. Assuming large (but finite) number of particles, 
we find to the leading order in $N$:
\[ B=1+\sigma^2N^{-1}+o(N^{-1}),\ B^N=e^{\sigma^2}+o(1).\]
Thus,
\begin{equation}
\frac{N}{N-1}f_2(r,q)\approx (1+\alpha +\alpha^2)\tilde{f}_1(r)\tilde{f}_1(q)+\alpha n^2
-2\alpha^2n^2\sigma^2e^{\sigma^2}N^{-1}\theta(r)\theta(q),
\end{equation}
where we have denoted $\theta=\sigma^{-1}\zeta$,  $\langle\theta^2\rangle=1$.
This correlation is negative once the sign of the deviations
from the homogeneity at points $r$ and $q$ are the same, and positive if these  
deviations have the opposite signs.

\section{Conclusion}
Once a classical statistical system is out of the thermodynamic limit, the exclusive character
of the Boltzmann-Gibbs-Shannon entropy  is fading away, and classical ensembles are not equivalent
anymore.  Whereas  using the microcanonical ensemble for any
description of  finite systems may be most appropriate,
this route is very complicated, at least  from a computational standpoint. 
For that reason, seeking an entropic description of effects
of finiteness is  a relevant option. 

We stress it once again, that the one-parametric family  $H_{\alpha}$, Eq.\  (\ref{result1}) and  (\ref{family3}),
is the {\it unique} generalization of the classical Boltzmann-Gibbs-Shannon entropy consistent
with the additivity and the trace-form requirements simultaneously.
It is reasonable therefore to study its applicability to a description of
statistical systems out of the strict thermodynamic limit. The main result
of this paper is the analytical description of the quasi-equilibria for this family of
the entropy functions. We have demonstrated  
that the solutions to  the entropy maximization problems are accessible
in a fairly simple way, and which amounts to studying a function of one
variable, ${\rm lm} a$. 
This makes studies of the non-classical ensembles
described herein relatively uncomplicated, especially in the vicinity
of the classical BGS solutions, where we expect, in the first place,
the theory to be meaningful. Eventually, predictions can be compared 
in molecular dynamics simulations by making the size of the system
smaller, and/or the number of particles smaller. This is left for a future work.

\bibliography{myrefs}

\end{document}